\documentclass[prd,aps,tightenlines,preprint,nofootinbib,superscriptaddress]{revtex4}
\usepackage{epsfig}
\usepackage{amsmath}
\input{epsf}

\newcommand{\nslash}{n \hspace{-0.22cm} / \,}

\begin{document}

\vspace*{2cm}
\title{Double spin asymmetry {\boldmath $A_{LT}$} in direct photon production}

\author{Zuo-Tang Liang}
\affiliation{School of Physics, Shandong University, Jinan, Shandong 250100, China}
\author{Andreas Metz}
\affiliation{Temple University, Philadelphia, PA 19122-6082, USA}
\author{Daniel Pitonyak}
\affiliation{Temple University, Philadelphia, PA 19122-6082, USA}
\author{Andreas Sch\"{a}fer}
\affiliation{ Institut f\"{u}r Theoretische Physik, Universit\"{a}t Regensburg, Regensburg, Germany}
\author{Yu-Kun Song}
\affiliation{Department of Modern Physics, University of Science and Technology of China, Hefei, Anhui 230026, China}
\author{Jian Zhou}
\affiliation{ Institut f\"{u}r Theoretische Physik, Universit\"{a}t Regensburg, Regensburg, Germany}

\date{\today}

\begin{abstract}
We study the longitudinal-transverse double spin asymmetry $A_{LT}$ for direct photon
production in nucleon-nucleon scattering by using the collinear twist-3 approach.
This asymmetry, which, for instance, could be measured at RHIC, contains a complete
set of collinear twist-3 correlation functions in a transversely polarized nucleon.
\end{abstract}

\maketitle

\section{Introduction}
High energy experiments with polarized beams and targets have opened a new window for revealing
QCD dynamics and hadron structure.
Based on QCD factorization theorems, polarization-dependent cross sections generally can be
factorized into the convolution of perturbatively calculable hard parts and universal
nonperturbative (soft) parts which are expressed through various spin-dependent parton
correlation functions.
Among these functions, higher-twist spin-dependent correlation functions are poorly known in
comparison to the three leading-twist ones: the spin-averaged parton distribution $f_1$,
the helicity distribution $g_1$, and the quark
transversity $h_1$~\cite{Ralston:1979ys,Cortes:1991ja,Jaffe:1991kp}.
Along with leading-twist distributions, the higher-twist correlation functions provide us with
important information on the structure of hadrons, even though they do not have a probability
interpretation.
The best way of extracting them is to investigate spin observables which have no leading-twist
contribution.
A classic example is the twist-3 double spin asymmetry $A_{LT}$ (longitudinally polarized lepton
beam, transversely polarized target) in inclusive deep-inelastic lepton-nucleon scattering (DIS),
which allows one to study the parton correlator $g_T$.
In the case of nucleon-nucleon scattering, the double spin asymmetry $A_{LT}$ in the Drell-Yan
process, involving two polarized incident hadrons, has been extensively
investigated for the purpose of studying higher-twist
correlators~\cite{Jaffe:1991kp,Tangerman:1994bb,Koike:2008du,Lu:2011th}.
More recently, $A_{LT}$ for inclusive lepton production from the decay of $W$-bosons in
proton-proton scattering and for jet production in lepton-proton scattering has been
derived~\cite{Metz:2010xs,Kang:2011jw}.
In this Letter we focus on the double spin asymmetry $A_{LT}$ for direct photon production in
nucleon-nucleon scattering, which also allows one to study twist-3 spin-dependent parton
distributions as leading effects.
In contrast to the aforementioned processes, direct photon production contains a complete set
of collinear twist-3 correlation functions in a transversely polarized nucleon.

The crucial tool required for the extraction of twist-3 correlation functions is collinear
higher-twist factorization, which has been established in studying transverse single spin
asymmetries (SSAs)~\cite{Efremov:1981sh,Qiu:1991pp,Qiu:1998ia,Eguchi:2006qz,Kouvaris:2006zy,Zhou:2009jm}.
The exploration of SSAs in hadronic reactions has a long history, starting from the mid 1970's.
In particular, the large size of the observed SSAs for single inclusive hadron
production~\cite{Bunce:1976yb,Adams:1991rw,Krueger:1998hz,Adams:2003fx,Adler:2005in,:2008mi} came
as a big surprise and, {\it a priori}, posed a challenge for QCD, because the collinear parton
model predicts the asymmetries are proportional to
$\alpha_s m_q / P_{h\perp}$~\cite{Kane:1978nd,Ma:2008gm}, where $m_q$ is the quark mass and
$P_{h\perp}$ is the transverse momentum of the final state hadron.
However, significant SSAs in hadronic collisions may be generated by going beyond the naive
parton model and including collinear twist-3 parton correlators, as was first pointed out
in~\cite{Efremov:1981sh}, and later on studied in more
detail~\cite{Qiu:1991pp,Qiu:1998ia,Eguchi:2006qz,Kouvaris:2006zy,Zhou:2009jm,Koike:2009ge}.
(We also note that alternative mechanisms underlying the large SSAs have been
proposed~\cite{Hoyer:2006hu,Qian:2011ya,Kovchegov:2012ga}.)
In the process $p^{\uparrow} p \to h X$, the collinear twist-3 formulation has some relation
to a description in terms of transverse momentum dependent parton correlators
(TMDs)~\cite{Sivers:1989cc,Collins:1992kk,Anselmino:1994tv,D'Alesio:2007jt}, provided that
initial/final state interactions in the TMD approach are taken into account~\cite{Gamberg:2010tj}.
In particular, for semi-inclusive DIS and related processes, TMDs are of crucial
importance (see for instance~\cite{Mulders:1995dh,Boer:1997nt,Brodsky:2002cx,Collins:2002kn,
Ji:2002aa,Liang:2006wp,Bacchetta:2006tn,Arnold:2008kf,Aybat:2011zv,Collins:book} and references
therein).
For these reactions, also intriguing nonzero spin/azimuthal asymmetries were
observed~\cite{Airapetian:2004tw,Alexakhin:2005iw,Avakian:2003pk,Abe:2005zx}.
In the recent past, important progress was made in understanding various spin observables
in terms of the collinear twist-3 approach and/or the TMD approach.
To mention just one example, it was found that for certain structure functions in
semi-inclusive DIS the two formalisms provide the same result at intermediate transverse hadron
momenta~\cite{Ji:2006ub,Bacchetta:2008xw,Zhou:2008fb,Yuan:2009dw}, which can be viewed as a
nontrivial consistency check.

By using the collinear higher-twist approach we are able to compute the pertinent spin-dependent
cross section (numerator of the asymmetry $A_{LT}$) for direct photon production in
nucleon-nucleon scattering and express it in terms of twist-3 parton distributions.
In Sect.~II, we briefly review the factorization formalism as it applies to the first
non-leading term of an expansion in powers of $1/l_{\gamma \perp}$, with $l_{\gamma \perp}$
denoting the transverse momentum of the observed photon.
We also list the twist-3 correlation functions that appear in the factorization formula and
provide the complete twist-3 spin-dependent cross section.
We conclude the paper in Sect.~III.

\section{Calculation of double spin dependent cross section}
For definiteness we consider the process
\begin{equation}
N(P,S_{\perp}) + N(\bar{P},\Lambda) \to \gamma(l_{\gamma}) + X \,,
\end{equation}
where we indicated the 4-momenta and the polarization states of the particles.
Furthermore, we define the Mandelstam variables through $S = (P + \bar{P})^2$,
$T = (P - l_{\gamma})^2$, and $U = (\bar{P} - l_{\gamma})^2$.
On the partonic level one has $\hat{s} = x x' S$, $\hat{t} = x T$, and $\hat{u} = x' U$,
with $x$ and $x'$ representing the longitudinal momentum fractions of active partons coming from
the transversely polarized nucleon and the longitudinally polarized nucleon, respectively.

\begin{figure}[t]
\begin{center}
\includegraphics[width=12cm]{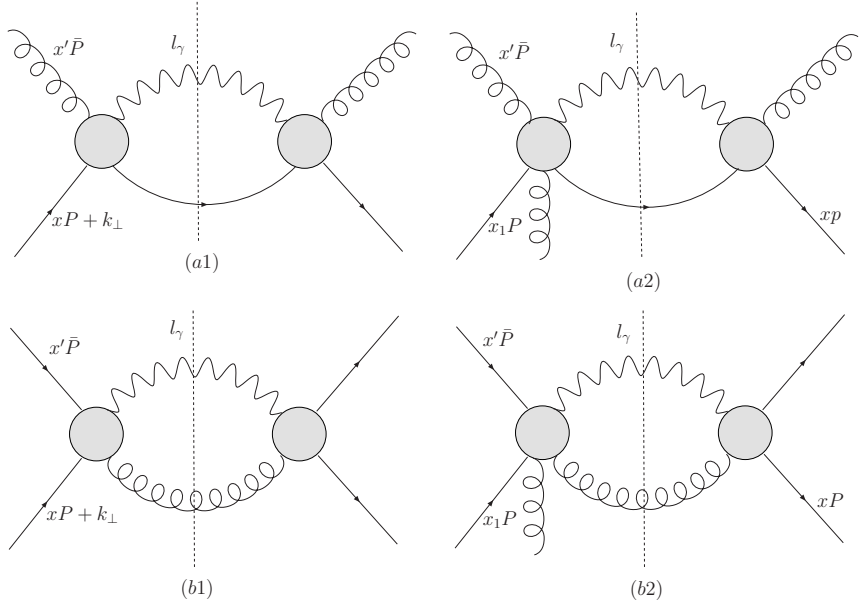}
\end{center}
\vspace{-0.4cm}
\caption{\it Generic Feynman diagrams for the partonic channels $qg \to q\gamma$ and
$q\bar{q} \to g\gamma$.
Diagrams (a1) and (b1) contribute to the hard part associated with the soft part
$\langle \bar \psi\partial_\perp \psi\rangle$;
diagrams (a2) and (b2) contribute to the hard part associated with the soft part
$\langle \bar \psi A_\perp \psi\rangle$.}
\label{f:diagrams}
\end{figure}

The generic form of the (spin-dependent) cross section reads
\begin{equation}
d \sigma( l_{\gamma \perp}, S_{\perp}, \Lambda ) =
H^{(0)} \otimes f_2 \otimes f_2 +
\frac{1}{l_{\gamma \perp}} H^{(1)} \otimes f_2 \otimes f_3
+ \mathcal{O} \bigg ( \frac{1}{l_{\gamma \perp}^2} \bigg) \,,
\label{e:gen_xs}
\end{equation}
where $f_2$ $(f_3)$ indicates a twist-2 (twist-3) parton distribution, while $H^{(0)}$ and
$H^{(1)}$ are the corresponding perturbatively calculable coefficient functions.
The direct products denote convolutions in the fractional parton momenta.
The second term in Eq.~(\ref{e:gen_xs}) gives rise to a leading non-vanishing contribution to
the double spin asymmetry $A_{LT}$.
The lowest-order contributions to $H^{(1)}$ are found from the Born diagrams for the partonic
process (see Figs.$\,$\ref{f:diagrams} (a1), (b1)), plus $\mathcal{O}(g)$ corrections in which
an extra gluon is exchanged between the remnants of the transversely polarized nucleon and the
hard partonic scattering process (see Figs.$\,$\ref{f:diagrams} (a2), (b2)).
In order to isolate the twist-3 contributions to the Born diagrams at this order, we employ a
collinear expansion in the parton momenta.

To proceed further, let us briefly review the relevant twist-3 correlation functions involved
in our calculation.
For the $A_{LT}$ asymmetry we consider twist-3 correlators in the transversely polarized nucleon,
together with ordinary twist-2 helicity distributions $g_1^a$ in the longitudinally polarized
nucleon.
The so-called D-type (twist-3) quark-gluon-quark functions (here denoted by $G_D$ and
$\tilde{G}_D$) have been introduced a long time ago and defined
through~\cite{Ellis:1982wd,Jaffe:1991kp}
\begin{eqnarray}
&& \int \frac{dy^-}{2 \pi} \frac{dy_1^-}{2\pi} \, e^{-ixP^+ y^-} \, e^{i(x-x_1)P^+y^-_1}
\langle P,S_\perp | \bar{\psi}_\beta(y^-)  i D_\perp^\mu(y_1^-)\psi_\alpha(0) | P,S_\perp \rangle
\nonumber \\
&& \hspace{0.5cm}
= \frac{M}{2P^+} \Big[ G_D(x,x_1) \, i\varepsilon^{\mu\nu}_\perp S_{\perp \nu} \nslash
+\tilde{G}_D(x,x_1) \, S_\perp^\mu \gamma_5 \nslash \Big]_{\alpha\beta} \,,
\label{e:D-type}
\end{eqnarray}
where the gauge links between the field operators have been suppressed.
The hadron momentum $P$ is proportional to the light cone vector $n = (1^+, 0^-,\vec{0}_{\perp})$,
whose conjugate light-cone vector is $\bar{n} = (0^+, 1^-, \vec{0}_{\perp})$.
The nucleon mass is denoted by $M$.
The gluon field enters through the covariant derivative
$D_{\perp}^{\mu} = \partial_{\perp}^{\mu} - i g A_{\perp}^{\mu}$.
The variables $x, x_1$ are the momentum fractions of the hadron carried by the quarks,
implying that the gluon momentum fraction is given by $x_g = x - x_1$.
We also recall the relation between the D-type functions in Eq.~(\ref{e:D-type}) and the twist-3
quark-quark correlator $g_T$~\cite{Jaffe:1991kp},
\begin{equation}
x g_T(x) = \int dx_1 \Big[ G_D(x,x_1) + \tilde{G}_D(x,x_1) \Big] \,.
\end{equation}
In the case of $A_{LT}$ in inclusive DIS, for instance, $G_D$ and $\tilde{G}_D$ appear with the
same hard scattering coefficient such that the final result is proportional to $g_T$.

Making use of the field strength tensor offers an alternative way of defining gauge invariant
quark-gluon-quark correlators~\cite{Efremov:1981sh,Qiu:1991pp},
\begin{eqnarray}
&& \int \frac{dy^-}{2 \pi} \, \frac{dy_1^-}{2\pi} \, e^{-ixP^+ y^-} e^{i(x-x_1)P^+y^-_1}
\langle P,S_\perp | \bar{\psi}_\beta(y^-)  g F_\perp^{+ \mu}(y_1^-)\psi_\alpha(0) | P,S_\perp \rangle
\nonumber \\
&& \hspace{0.5cm}
= \frac{M}{2} \Big[ T_F(x,x_1) \, \varepsilon^{\nu\mu}_\perp S_{\perp \nu} \nslash
+ \tilde{T}_F(x,x_1) \, i S_\perp^\mu \gamma_5 \nslash \Big]_{\alpha\beta} \,,
\end{eqnarray}
where $T_F$ and $\tilde{T}_F$ are the so-called F-type functions.
Note that our definition of these functions differs by a factor $2 \pi M$ from the
conventions used in Ref.~\cite{Ji:2006ub,Zhou:2008fb}.
It was found that the F-type functions directly enter the QCD-description of transverse
SSAs in various processes~\cite{Qiu:1991pp,Qiu:1998ia,Eguchi:2006qz,Kouvaris:2006zy}.

The D-type functions and the F-type functions are not independent as they can be related
to each other by means of the equation of motion~\cite{Boer:2003cm,Eguchi:2006qz},
\begin{eqnarray}
&& G_D(x,x_1) = P \frac{1}{x-x_1} \, T_F(x,x_1) \,,
\label{e:DF_1}
\\
&& \tilde{G}_D(x,x_1)= P \frac{1}{x-x_1} \, \tilde{T}_F(x,x_1) + \delta(x-x_1) \, \tilde{g}(x) \,,
\label{e:DF_2}
\end{eqnarray}
where $P$ indicates the principal value prescription, and the function $\tilde{g}$ is given
by~\cite{Boer:2003cm,Eguchi:2006qz}
\begin{eqnarray}
&& \int \frac{dy^-}{2 \pi} \, e^{-ixP^+ y^-}
\langle P,S_\perp | \bar{\psi}_\beta(y^-)
\bigg( iD_\perp^\mu + g\int_0^\infty d\zeta^- F_\perp^{+\mu}(\zeta^-) \bigg)
\psi_\alpha(0) | P,S_\perp \rangle
\nonumber\\
&& \hspace{0.5cm}
= \frac{M}{2} \Big[ \tilde{g}(x) \, S_\perp^\mu \gamma_5 \nslash \Big]_{\alpha\beta} \,.
\end{eqnarray}
Because of the relations in Eqs.~(\ref{e:DF_1}), (\ref{e:DF_2}) we may either use the D-type
or the F-type functions for our calculation.
However, it is mandatory to also include $\tilde{g}$ in order to completely describe
the spin-dependent cross section.
(See also~\cite{Tangerman:1994bb,Belitsky:1997zw,Kundu:2001pk,Metz:2008ib,Accardi:2009au}
and references therein for a discussion about independent twist-3 correlation functions.)
It is worthwhile to point out that the function $\tilde{g}$ is related to the twist-2
TMD $g_{1T}$ describing the distribution of longitudinally polarized quarks
in a transversely polarized nucleon~\cite{Zhou:2008mz,Zhou:2009jm},
\begin{equation}
\tilde{g}(x) = \int d^2 \vec{k}_{\perp} \, \frac{\vec{k}_{\perp}^{\,2}}{2 M^2} \,
g_{1T}(x,\vec{k}_{\perp}^{\,2}) \,,
\end{equation}
where we used the definition of $g_{1T}$ as given in~\cite{Mulders:1995dh,Bacchetta:2006tn}.

As mentioned above, the twist expansion is the key step in our calculations.
The relevant technical ingredients required for such a twist-3 analysis have been well
developed in the last few
decades~\cite{Efremov:1981sh,Ellis:1982wd,Qiu:1991pp,Qiu:1998ia,Eguchi:2006qz,Kouvaris:2006zy,Yuan:2009dw}.
In the twist expansion, a set of non-perturbative matrix elements of the hadron state is analyzed
according to the power counting of the associated contributions.
At the twist-3 level, the following matrix elements can contribute~\cite{Ellis:1982wd}:
\begin{equation}
\langle \bar \psi\partial_\perp \psi\rangle, \qquad
\langle \bar \psi A_\perp \psi\rangle, \qquad
\langle \bar \psi\partial_\perp A^+ \psi\rangle \,.
\end{equation}
We found it most convenient to work in the light-cone gauge ($A^+ = 0$) in which only the
first two matrix elements survive.
The matrix element $\langle \bar \psi\partial_\perp \psi\rangle$ can be transformed into the
gauge invariant matrix element $\tilde{g}$.
Moreover, by partial integration, the matrix element $\langle \bar \psi A_\perp \psi\rangle$
can be expressed as $\langle \bar \psi \partial^+ A_\perp \psi\rangle$ and further be related
to the gauge invariant matrix elements $T_F$ and $\tilde{T}_F$.
These functions may then be rewritten in terms of the D-type functions by means of
Eqs.~(\ref{e:DF_1}), (\ref{e:DF_2}).

Our goal is to perturbatively calculate the hard scattering coefficients associated with these
two soft parts.
The corresponding partonic scattering processes are illustrated in Fig.$\,$\ref{f:diagrams}.
It turns out that the hard parts are not always real.
Imaginary contributions occur whenever an internal parton line in the hard scattering goes
on-shell.
When this happens, we use the distribution identity
\begin{equation}
\frac{1}{x \pm i\epsilon}=P \frac{1}{x} \mp i \pi \delta(x) \,.
\end{equation}
While for the related calculations of transverse SSAs only the $\delta$-function contribution
matters, in this calculation for $A_{LT}$ we are left with the principal value part.
To be more precise, the imaginary parts always cancel between the different cut diagrams.
In order to combine corresponding cut diagrams we use the symmetry properties
\begin{eqnarray}
T_F(x,x_1) & = & T_F(x_1,x) \,, \hspace{0.95cm} \tilde{T}_F(x,x_1)=-\tilde{T}_F(x_1,x) \,,
\nonumber \\
G_D(x,x_1)& = & -G_D(x_1,x) \,, \quad  \tilde{G}_D(x,x_1)=\tilde{G}_D(x_1,x)
\end{eqnarray}
of the F-type functions and the D-type functions.

Making use of these ingredients, the calculation is rather straightforward, and we obtain the
following result for the spin-dependent cross section:
\begin{eqnarray}
&& \hspace{-1.0cm} l_{\gamma}^0 \frac{d\sigma(S_\perp, \Lambda)}{d^3 l_\gamma} =
-2 M \frac{\alpha_s \alpha_{em}}{S} \, \Lambda \, \vec{S}_\perp \cdot \vec{l}_{\gamma \perp}
\sum_{i = qg,q\bar{q}} \; \sum_{a,b} e_a^2 \int^1_{x'_{min}} \frac{dx'}{x'} \,
\frac{1}{x'S+T} \, \frac{1}{x \hat{u}} \, g_{1}^{b}(x')
\nonumber \\
&& \times \bigg\{ \Big[ \tilde{g}^{a} (x) -x \frac{d}{dx}\tilde{g}^{a} (x) \Big] H_{i}^{\tilde{g}}
+\int_0^1 dx_1 \Big[ G_{D}^{a}(x,x_1) H_{i}^{G_D} + \tilde{G}_{D}^{a}(x,x_1) H_{i}^{\tilde{G}_D}
\Big] \bigg\} \,,
\end{eqnarray}
where $x = - x'U / (x'S + T)$ and $x'_{min} = - T / (S + U)$.
The hard coefficient functions for the $qg \to q \gamma$ partonic channel are
\begin{eqnarray}
H^{\tilde{g}}_{qg} & = &  \frac{N_c}{N_c^2 -1}
\bigg[ \frac{\hat{s}^2 - \hat{t}^2}{\hat{s} \, \hat{t}} \bigg] \,,
\nonumber \\
H^{G_D}_{qg} & = &\frac{N_c}{N_c^2-1}
\bigg[ \frac{\hat{s}^2 - \hat{t}^2}{(1 - \xi) \, \hat{s} \, \hat{t}} \bigg]
+ \frac{1}{N_c} \bigg[ \frac{\hat{u} \, (\hat{s}^2 + 2\hat{t}^2)}{\hat{s} \, \hat{t}^2}
+ \frac{\hat{u}}{(1-\xi) \, \hat{t}} \bigg] \,,
\nonumber \\
H^{\tilde{G}_D}_{qg} & = & \frac{N_c}{N_c^2-1}
\bigg[ \frac{(\xi - 2) \, (\hat{s}^2 - \hat{t}^2)}{\xi (1- \xi) \, \hat{s} \, \hat{t}} \bigg]
+ \frac{1}{N_c} \bigg[ \frac{\hat{u} \, (\hat{s}^2 + 2\hat{t}^2)}{\hat{s} \, \hat{t}^2}
+ \frac{(\xi-2) \, \hat{u}}{\xi (1-\xi) \, \hat{t}}
+ \frac{2 \hat{u}}{\xi \, \hat{s}} \bigg] \,,
\end{eqnarray}
while the coefficient functions for the $q \bar{q} \to g \gamma$ channel read
\begin{eqnarray}
H^{\tilde{g}}_{q\bar{q}}& = &  \frac{1}{N_c^2 }
\bigg[ \frac{\hat{t}^2 + \hat{u}^2}{\hat{t} \, \hat{u}} \bigg] \,,
\nonumber \\
H^{G_D}_{q\bar{q}} & = &
 \frac{\hat{t}^2+\hat{u}^2}{(1 - \xi) \, \hat{t} \, \hat{u}}
+\frac{2C_F}{N_c} \bigg[ \frac{\hat{s}^2 \, (\hat{t} - \hat{u})}{\hat{t}^2 \, \hat{u}}
- \frac{(\xi - 2) \, (\hat{t} - \hat{u})}{(1 - \xi) \, \hat{t}} \bigg] \,,
\nonumber \\
H^{\tilde{G}_D}_{q\bar{q}} & = &
 \frac{(\xi - 2) \, (\hat{t}^2 + \hat{u}^2)}{\xi (1 - \xi) \, \hat{t} \, \hat{u}}
+ \frac{2C_F}{N_c} \bigg[ \frac{\hat{s}^2 \, (\hat{t} - \hat{u})}{\hat{t}^2 \, \hat{u}}
- \frac{\xi \, (\hat{t} - \hat{u})}{(1 - \xi) \, \hat{t}} \bigg] \,,
\end{eqnarray}
with $\xi=x_g/x$.
Note that the spin-dependent cross section relevant for the $A_{LT}$ asymmetry is
characterized by the correlation $\Lambda \, \vec{S}_\perp \cdot \vec l_{\gamma \perp}$
instead of the correlation $\vec{S}_\perp \times \vec l_{\gamma \perp}$ that arises in
the case of transverse SSAs.
We have expressed the cross section in terms of the D-type functions plus $\tilde{g}$.
By doing so the contributions from the derivative term and the non-derivative term
associated with the correlator $\tilde{g}$ can be combined into the same compact form
that was found for the SSAs for direct photon production and inclusive pion
production~\cite{Qiu:1991pp,Qiu:1998ia,Kouvaris:2006zy}.
We also point out that, unlike related previous studies of other
processes~\cite{Metz:2010xs,Kang:2011jw}, the coefficient functions for $G_D$ and for
$\tilde{G}_D$ are different.
Therefore, in the present case it does not pay off to involve $g_T$ in the final result.
This also means that through direct photon production, in combination with other reactions,
one should in principle be able to study a compete set of twist-3 correlators for a
transversely polarized nucleon.

\section{Summary}
By using collinear twist-3 factorization we derived the longitudinal-transverse double
spin asymmetry $A_{LT}$ for direct photon production in nucleon-nucleon scattering.
Our study can be considered as the counterpart of the calculation of the transverse SSA
in the same process~\cite{Qiu:1991pp}.
Measuring this observable might also open a new window to test the higher-twist approach,
for instance by looking at the transverse momentum behavior of $A_{LT}$.
Also, due to the derivative term of the correlator $\tilde{g}$, the asymmetry may be largest
in the large $x_F$ region --- the same kinematics for which the largest SSAs in hadron
production have been observed.
Moreover, if the gluon helicity distribution (in the small $x$ region) is very small
(see, e.g., Ref.~\cite{deFlorian:2008mr}), one can expect $A_{LT}$ to be larger for
$p\bar{p}$ scattering than for $pp$-scattering.

The result for $A_{LT}$ depends on a complete set of collinear twist-3 parton correlators
for a transversely polarized nucleon.
It requires numerical studies to find out to what extent these correlators could, at least
in principle, be separately explored by variation of the kinematics.
Most likely one would need the combined information from different processes to address
all three twist-3 correlation functions.
Besides the twist-3 effect investigated in this paper, a collinear twist-3 correlator for the
longitudinally polarized hadron, coupling to the transversity distribution, may also contribute
to $A_{LT}$.
This part as well as numerical estimates are left for future study.
One can further extend our work to the $A_{LT}$ asymmetry in other processes such as single
inclusive hadron production and jet production~\cite{Daniel}.
\\[0.3cm]
{\bf Acknowledgement:} One of us (J.Z.) thanks Feng Yuan for helpful discussions and encouragement.
This work has been supported in part by the BMBF (OR 06RY9191),
the National Science Foundation (PHY-0855501), and the National Natural Science Foundation
of China (10975092 and 11035003).


\end {document}